\documentclass[structabstract]{aa}  
\usepackage{graphicx}
\usepackage{txfonts}
\begin{document}

   \title{Rotation-stimulated structures in the CN and C$_3$ comae of comet 103P/Hartley 2 around the {\em EPOXI} 
    encounter\thanks{Based on data collected with \mbox{2-m} RCC telescope at Rozhen National Astronomical Observatory}}


   \author{W. Waniak\inst{1}
          \and
          G. Borisov\inst{2}
          \and
          M. Drahus\inst{3}
          \and
          T. Bonev\inst{4}}

   \institute{Astronomical Observatory of the Jagiellonian University, Orla 171, 30-244 Krakow\\
              \email{waclaw.waniak@uj.edu.pl}
         \and
             Institute of Astronomy of the Bulgarian Academy of Sciences, 72 Tsarigradsko Chaussee Blvd., 1784 Sofia\\
             \email{gborisov@astro.bas.bg}
         \and
             University of California at Los Angeles, 595 Charles E. Young Dr. East, Los Angeles, CA 90024\\
             \email{mdrahus@ucla.edu}
         \and
             Institute of Astronomy of the Bulgarian Academy of Sciences, 72 Tsarigradsko Chaussee Blvd., 1784 Sofia\\
             \email{tbonev@astro.bas.bg}}

   \date{Received ; accepted}

 
  \abstract
   {In late 2010 a Jupiter Family comet 103P/Hartley 2 was a subject of an intensive \mbox{world-wide} investigation.
    On UT October 20.7 the comet approached the Earth within only 0.12 AU, and on UT November 4.6 it was visited
    by NASA \begin{em} EPOXI \end{em} spacecraft.}
   {We joined this international effort and organized an observing campaign with three key goals.
    First, to measure the parameters of the nucleus rotation in a time series of CN. Second, to investigate
    the compositional structure of the coma by comparing the CN images with nightly snapshots of C$_3$. And third,
    to investigate the photochemical relation of CN and HCN, using the HCN data collected
    nearly simultaneously with our images.}
   {The images were obtained through narrowband filters using the \mbox{2-m} telescope of the Rozhen National Astronomical
    Observatory. They were taken during 4 nights around the moment of the \begin{em} EPOXI \end{em} encounter.
    Image processing methods and periodicity analysis techniques were used to reveal transient coma
    structures and investigate their repeatability and kinematics.}
   {We observe shells, arc-, jet- and \mbox{spiral-like} patterns, very similar for the CN and C$_3$ comae. The CN features
    expanded outwards with the \mbox{sky-plane} projected velocities between 0.1 to 0.3 \mbox{km s$^{-1}$}.
    A corkscrew structure, observed on November 6, evolved with a much higher velocity of 0.66 \mbox{km s$^{-1}$}.
    Photometry of the inner coma of CN shows variability with a period of \mbox{18.32$\pm$0.30 h} (valid for
    the middle moment of our run, UT 2010 Nov. 5.0835), which we attribute to the nucleus rotation.
    This result is fully consistent with independent determinations around the same time by other teams.
    The pattern of repeatability is, however, not perfect, which is understendable given the
    suggested excitation of the rotation state, and the variability detected in CN correlates well with the cyclic
    changes in HCN, but only in the active phases. The revealed coma structures, along with the snapshot of the
    nucleus orientation obtained by \begin{em}EPOXI\end{em}, let us estimate the spin axis orientation.
    We obtained \mbox{RA=122\degr}, \mbox{Dec=+16\degr} (epoch J2000.0), neglecting at this point the rotational
    excitation.}
   {}


   \titlerunning{Rotation-stimulated structures in 103P/Hartley 2}
   \authorrunning{Waniak et al.}

   \maketitle
%

\section{Introduction}

   103P/Hartley 2 (hereafter 103P) is a Jupiter Family comet, which currently has a perihelion at 1.06 AU
   and orbits the Sun with a \mbox{6.47-year} period. Due to frequent perihelion passages, the comet was relatively
   well characterised prior the 2010 apparition (see e.g. Arpigny at al. \cite{arpigny93}; Weaver et al. \cite{weaver94};
   Crovisier et al. \cite{crovisier99}; Colangeli et al. \cite{colangeli99}; Epifani et al. \cite{epifani01}),
   although no information about the nucleus rotation was obtained.

   The state of rotation is, however, of great interest, because it is one of the parameters which establish
   the lifetime of a comet nucleus. It has been realised long ago, that a nucleus rotating faster than a certain
   limit (see e.g. Davidsson \cite{davidsson01}) must be disrupted by a centrifugal force. On the other hand, knowledge of
   this parameter enables proper interpretation of the observational data and their conversion to the physical quantities
   (such as the total production rates and relative abundances) which can be compared among comets.
 
   The rotation period of 103P was first measured at \mbox{16.4$\pm$0.2 h} (Meech et al. \cite{meech11}) shortly
   after the comet was selected as the target of NASA {\em EPOXI} mission. The studies on the rotation continued when
   the comet became active.
   \mbox{{\em EPOXI}-based} imaging and photometry (A'Hearn et al. \cite{ahearn11}), as well as \mbox{ground-based} observations
   (see Meech et al. \cite{meech11} for an overview of the \mbox{ground-based} results), show that the rotation period was increasing
   with time and that the nucleus most probably was rotating in excited mode.

   Although indications of possible changes in the spin rate among comets have been reported before for some
   objects (e.g. \mbox{C/1990 K1} (Levy), \mbox{C/2001 K5} (LINEAR), \mbox{2P/Encke}, \mbox{6P/d'Arrest}, \mbox{10P/Tempel 1}),
   the first unequivocal measurement of a slow decrease of the rotation period was obtained only recently for comet \mbox{9P/Tempel 1}
   (Belton \& Drahus \cite{belton07}; Belton et al. \cite{belton11}). Model computations
   (e.g. Guti\'errez et al. \cite{gutierrez03})
   show that emission of matter from active vents occupying small fraction of an \mbox{irregularly-shaped} rotating
   nucleus can produce a significant net torque. Due to this effect cometary
   rotation can evolve from the \mbox{principal-axis} toward the \mbox{non-principal-axis}
   (excited) mode after a couple of perihelion passages and remain in this state during the next tens of orbital
   revolutions.
   The problem of rotational excitation is important for the \mbox{long-term} evolution of cometary nuclei as the
   relaxation timescale has not been well established to date (e.g. Samarasinha et al. \cite{samarasinha04}).
   Despite the suggestion made by some authors (e.g. Belton \cite{belton91a}; Jewitt \cite{jewitt92}) that most
   \mbox{short-period} comets should be in excited spin states, only a couple of such cases have been reported before:
   \mbox{2P/Encke} (Belton et al. \cite{belton05}),
   \mbox{29P/Schwassmann-Wachmann 1} (Meech et al. \cite{meech93}), and \mbox{1P/Halley}
   (Belton et al. \cite{belton91b}). The recent results on 103P show that
   its nucleus -- having been observed in an excited and decelerated rotation state -- is the first known 
   comet in which the two processes occur at the same time. Moreover, thanks to the {\em EPOXI}
   mission, which provided detailed characteristics of the nucleus, this exceptional object is a unique laboratory
   to test our theories about the rotational dynamics of cometary nuclei.

   Impressive images of 103P's nucleus obtained by {\em EPOXI} (A'Hearn et al. \cite{ahearn11})
   show an elongated body with tips of different
   sizes, the smaller one being insolated and active at the encounter. The nucleus is found to precess
   about the longest axis of inertia with the period of
   \mbox{18.34$\pm$0.04 h} and roll about the shortest axis of inertia (longest nucleus extent) with the probable
   period of
   \mbox{27.79$\pm$0.31 h} (both at the epoch of the {\em EPOXI} encounter). The two periodicities are nearly commensurate
   in 2:3 resonance which causes the activity pattern to repeat every three precession cycles -- the
   characteristic observed also in \mbox{ground-based} HCN data (Drahus et al. \cite{drahus11}).

   In this work we analyse transient features in the CN and C$_3$ comae. In particular,
   the time series of CN is used to examine the kinematics, repeatability, and periodicity in the framework
   of the nucleus \mbox{precession-roll} scenario. We also determine the spin axis orientation of the nucleus.
   Our campaign was realized simultaneously with the HCN monitoring at IRAM \mbox{30-m} (Drahus et al. \cite{drahus11};
   \cite{drahus12}), which makes it possible to investigate the connection between HCN and CN.
   Although the HCN molecule is considered
   as the main donor of cometary CN (e.g. \mbox{Bockel\'ee-Morvan} \& Croviser \cite{bockelee85}),
   the issue is still somewhat controversial (e.g. Fray et al. \cite{fray05}).

\section{Observations and data reduction}

   Observations of 103P were carried out on four photometric nights: November 2, 4, 5 and 6, 2010, including
   the date (though not the moment) of the {\em EPOXI} encounter. The last night before the encounter, i.e.
   November 3, 2010, was also allocated but lost due to bad weather. We observed with the \mbox{two-channel} FoReRo2
   focal reducer (Jockers et al. \cite{jockers00}) mounted in the Cassegrain focus of the \mbox{2-m} 
   \mbox{Ritchey-Chr\'etien-Coud\'e}
   telescope of the Rozhen National Astronomical Observatory (Bulgarian Academy of Sciences). In the blue channel we used
   Photometrics \mbox{CE200A-SITe}, and in the red channel VersArray 512B. Both CCDs have
   \mbox{square-shaped} \mbox{24-$\mu$m} pixels, and give the image scale of 0.89 arcsec pix$^{-1}$, i.e. \mbox{$\sim$100} km at the comet
   nucleus.

   In this work we present the results of imaging in the blue channel and the other data will
   be presented in a subsequent paper. Emission bands of CN and C$_3$ were observed through the HB filters
   (Farnham et al. \cite{farnham00}) at 387 nm and 406 nm respectively; dust was observed through an interference filter
   at 443 nm. All exposures were 600 s with \mbox{non-sidereal} tracking on the comet. We obtained a couple of hours
   long series of CN images and snapshots of C$_3$ and dust. Reduction of the images included, in addition to the
   standard steps, removal of Cosmic Ray Hits (CRH) and stellar profiles. We stacked consecutive images in pairs
   and cleaned them using the procedure described
   in Appendix A. In this way we obtained 31 clean stacks for CN, 7 for C$_3$ and 7 for dust. 
   The \mbox{mid-exposure} moments for the \mbox{co-added} images have been corrected for the travel time of light.
   The statistics of the observations and the geometric circumstances, are presented in Table 1. 


\begin{table}
\caption{Statistics of the images and geometric circumstances}             
\label{table:1}      
\centering                          
\begin{tabular}{c c c c c}        
\hline \hline                
\noalign{\smallskip}
UT Date  & $r$ $^{\mathrm{a}}$ & $\Delta$ $^{\mathrm{b}}$ & $\phi$  $^{\mathrm{c}}$ & Number of exposures $^{\mathrm{d}}$ \\    
Nov. 2010  & [AU]  & [AU] & [$^\circ$]  & CN\,\,\,\,\,\,\,\,\,C$_3$\,\,\,\,\,\,dust  \\
\noalign{\smallskip}
\hline                        
\noalign{\smallskip}
   2.96 -- 3.10 & 1.062 & 0.150 & 58.7 & \,\,\,\,11/6 \,\,\,\,\,\,2/1 \,\,\,\,4/2 \,\,\,\,\, \\      
   4.94 -- 5.16 & 1.064 & 0.158 & 58.8 & \,\,\,\,20/10 \,\,\,4/2 \,\,\,\,\,2/1 \,\,\,\,\, \\      
   5.93 -- 6.16 & 1.066 & 0.163 & 58.8 & \,\,\,\,17/9 \,\,\,\,\,\,3/2 \,\,\,\,\,3/2 \,\,\,\,\, \\      
   6.96 -- 7.17 & 1.067 & 0.167 & 58.7 & \,\,\,\,11/6 \,\,\,\,\,\,3/2 \,\,\,\,\,3/2 \,\,\,\,\, \\      
\noalign{\smallskip}
\hline             
\end{tabular}

\begin{list}{}{}
\item[$^{\mathrm{a}}$] Heliocentric distance 
\item[$^{\mathrm{b}}$] Topocentric distance 
\item[$^{\mathrm{c}}$] Topocentric phase angle (i.e. \mbox{Sun-comet-observer} angle) 
\item[$^{\mathrm{d}}$] Numbers of original/stacked images.   
\end{list}
\end{table}

   To enhance the visibility of faint coma structures we proceeded in a similar way as for comet 8P/Tuttle
   (Waniak et al. \cite{waniak09}). This approach assumes and critically depends on constant heliocentric distance
   of the comet and constant observing geometry in the analysed set of images -- the conditions which are
   well satisfied for our short run (see Table 1). In such a case, the only reason for coma
   variability can be transient phenomena produced by e.g. unstable and/or periodically
   changing mass ejection. First, we converted the scales of the individual stacked images to the same topocentric
   distance for UT 2010 Nov. 5.0. Then a series of the rescaled frames for a given filter was used as an input to
   our novel technique: the {\em Iterative Image Decomposition}, which extracts the \mbox{time-invariant} coma
   profile, and produces a series of images for the residual, \mbox{time-dependent} component. Our iteration loop
   contains two steps:
   ({\em i}) rescaling image signal by a factor followed by ({\em ii}) stacking such modified images into a mean
   frame.
   The normalization factors are adjusted with respect to this mean frame taken as a reference. The procedure
   begins with stacking of the original input frames with no normalization (i.e. the factors set equal to unity).
   In the subsequent iterations we stacked not the whole images but only the parts that do not change their profiles
   from frame to frame. These stable parts made it possible to comput the factors for signal rescaling.
   To detect and avoid pixels containing signatures of temporary signal enhancement we applied the $K\sigma$
   criterion for the pixel value difference between the individual normalised frame and the mean image. 
   After a number of iterations, when the mean profile appeared stable, we took this profile as the
   \mbox{time-invariant} coma pattern
   and subtracted it from the individual, normalised input frames, obtaining a series of {\em residual images}.
   Our experience shows, that such images are photometrically linked to a much better precision than can be
   achieved when linking through standard stars and nightly extinction coefficients. This
   approach has also
   other advantages over typical image enhancement procedures. It generally preserves photometric information,
   so the transient structures can be analysed quantitatively. Moreover, the level of enhancement does not depend
   on position, shape or size of a feature, if only the area of this feature is significantly smaller than the
   total area of the analysed part of the coma, and if the series of input images sufficiently samples the time
   variability of the coma profile.
\begin{figure}
  \centering
  \includegraphics[width=80mm]{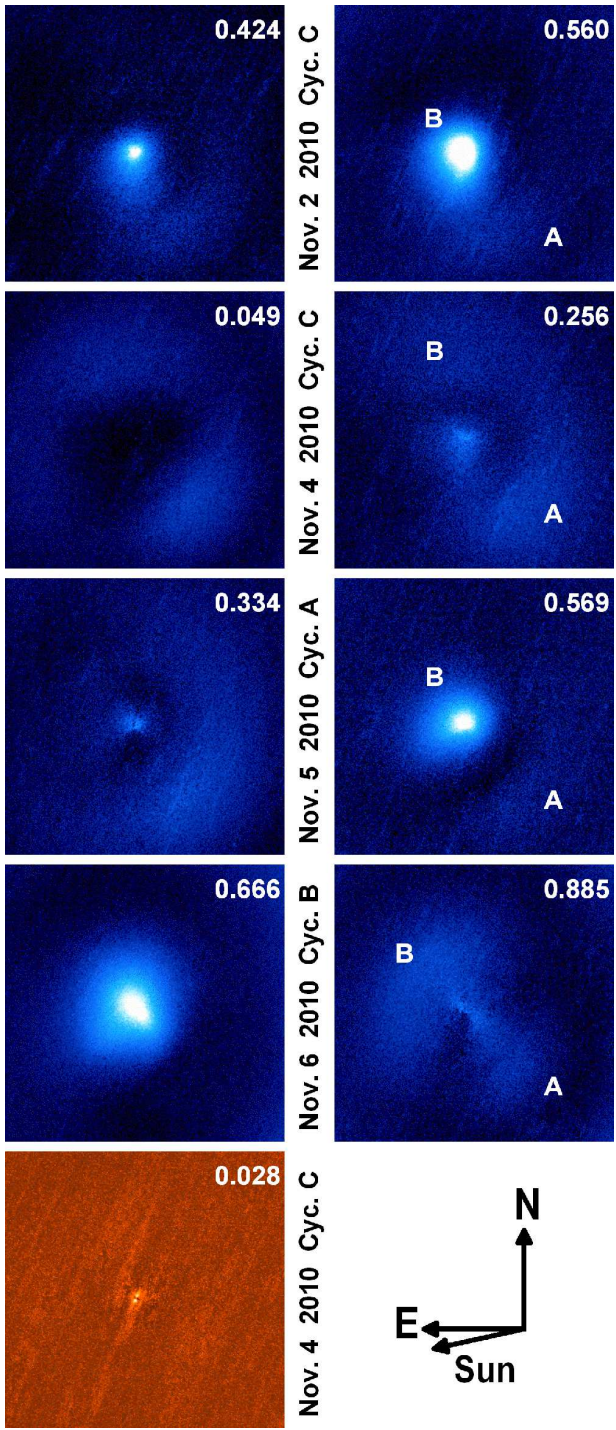}
  \caption{First and last residual image of CN for each night.
           The bottom left panel shows the typical residual image of the dust coma.
           Dates and cycles (see text for details) are displayed.
           Phases with respect to the \mbox{18.32 h} cycle are shown in the \mbox{upper-right} corners.
           The field of view is 6.4 arcmin or \mbox{43.8$\times$10$^{3}$} km at the nucleus.}
  \label{Onef}
\end{figure}
   As we did not subtract the light scattered by cometary dust from the signal in the molecular filters,
   it is obvious that our CN and C$_3$ images still contain dust contribution. This contribution is very weak,
   though, as 103P has a very low \mbox{dust-to-gas} ratio (Schleicher \cite{schleicher10}). Furthermore, the dust images
   processed with our {\em Iterative Image Decomposition}, show no transient structures (see Fig.~\ref{Onef}).
   This means that the signal contribution from dust is totally contained in the stationary CN and C$_3$ profiles,
   and therefore the {\em residual images} contain pure information about the variability of the molecular coma.

\section{CN and C$_3$ transient structures}

\begin{figure}
  \centering
  \includegraphics[width=84mm]{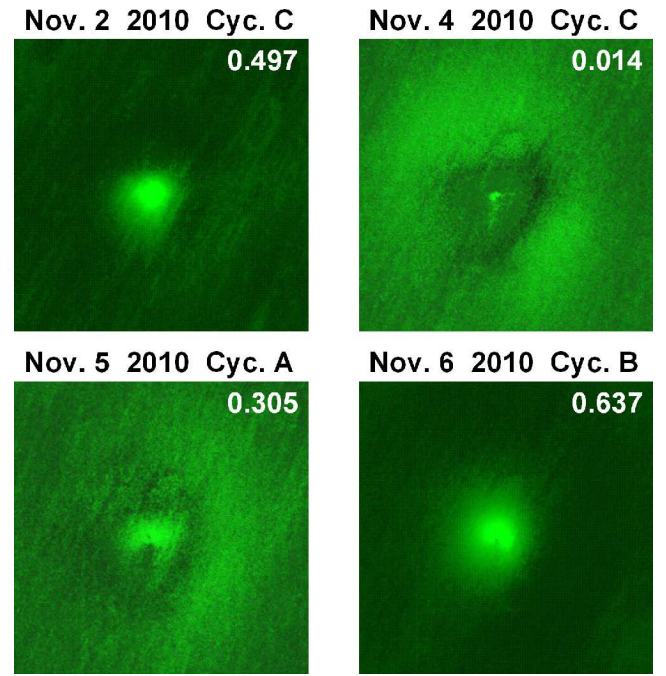}
  \caption{Images of C$_3$ obtained by stacking the nightly series of the residual frames
           to increase the $S/N$ ratio. Dates and cycles (see text for details)
           are displayed above the frames. Phases with respect to the \mbox{18.32 h} period are
           shown in the \mbox{upper-right} corners. Orientation and image scale are the same as in Fig.~\ref{Onef}.} 
  \label{Twof}
\end{figure}
   Our {\em residual images} for CN and C$_3$ show the evolution of the transient structures during 4 nights
   between November 2 and November 6.
   Part of the whole set for CN is presented in Fig.~\ref{Onef} (the complete, chronologically ordered series can be seen as
   a video clip Animation1). To show how the CN coma evolved with time, in Fig.~\ref{Onef} we present the first (left column)
   and last (right column) {\em residual image} for each night. In Fig.~\ref{Twof} we show \mbox{nightly-averaged} snapshots for C$_3$.
  
   Figures \ref{Onef} and \ref{Twof} reveal transient features of different kinds, and show that the CN and C$_3$ structures
   fully correlate. This means that both molecular environments behaved in very similar ways. To analyse
   the coma evolution with time we introduced the nomenclature used earlier by Drahus et al. (\cite{drahus11}). Each
   {\em residual frame} corresponds to a given precession phase (changing from 0 to 1) and precession cycle
   (denoted as {\em A}, {\em B}, and {\em C}) of the threefold precession period (the \mbox{{\em three-cycle}} period).
   The phases are obtained using the precession period of \mbox{18.32 h} (justified further in Section 5). The moment of the
   {\em EPOXI} encounter occurred at the precession phase of 0.5 at the middle of {\em Cycle B}.

   As can be seen in Fig.~\ref{Onef}, our data cover almost the full precession cycle. During the four nights we observed
   {\em Cycles} {\em C}, {\em C}, {\em A} and {\em B}, respectively. Although we covered all the three cycles,
   we never observed the same cycle and phase twice. In general, we notice
   markedly enhanced CN production at phases close to 0.5 on November 2, 5 and 6, when the central shells appeared
   and brightened on November 2 and 5, and diminished on November 6. These shells expanded
   outward, bearing \mbox{arc- and spiral-like} structures, which are visible up to the early phases of the next
   cycles. The shape and temporal evolution of the structure from November 6 resembles a traditional 
   corkscrew (cf. Fig.~\ref{Onef} and Animation1): the \mbox{north-east} part can be identified
   as a relatively massive handgrip and the narrower \mbox{south-west} pattern can be
   attributed to the spiral (although its actual spirality is not visible due to projection onto the sky).    
   Such corkscrew patterns have also been reported for 103P earlier (Knight \& Schleicher \cite{knight11};
   Samarasinha et al. \cite{samarasinha11}). 

   Even a simple analysis reveals that all the three central shells have different profiles and 
   behave differently in time. This can can be seen in video clip Animation2, where we applied a
   technique similar to
   \mbox{{\em ring-masking}} (A'Hearn et al. \cite{ahearn86}) to the {\em residual images}, to additionally
   enhance the contrast in the central region (from the individual pixel values our method subtracts the smoothed
   radial profile of the azimuthal minimum). The
   features in the outer coma poorly correlate with their progenitors visible during the earlier cycles,
   which can be caused by the postulated excitation of the rotation state.

\section{Kinematics of the CN features}

\begin{figure}
  \resizebox{\hsize}{!}{\includegraphics{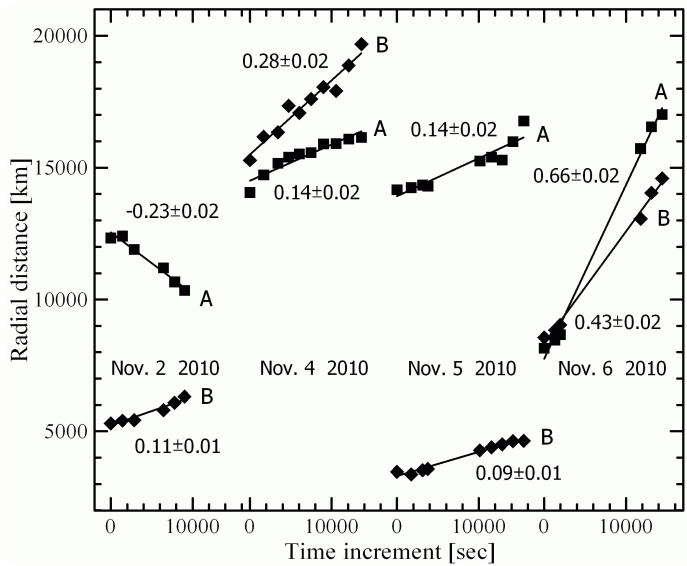}}
  \caption{Relation between the radial distances of the CN structure and time.
           The error bars are of the order of the symbol size and have been omitted.
           Linear fits and the expansion velocities in \mbox{km s$^{-1}$} are displayed.}
  \label{Threef}
\end{figure}

   To shed more light on the nature of the transient structures we have investigated their kinematics.
   The expansion
   velocity of the shells and arcs was measured separately in the \mbox{south-west} ({\bf A}) and \mbox{north-east} ({\bf B})
   quadrants, as labeled in Fig.~\ref{Onef}. We proceeded in a similar way as in our earlier work on 8P/Tuttle (Waniak et al.
   \cite{waniak09}). First, we transformed the CN {\em residual images} to the rectangular frames
   arranging the radial distance from the nucleus in columns and the azimuth angle in rows. Then
   for every pair of such images from a given night, we determined the radial shift of the transient
   feature between both images. We \mbox{cross-correlated} the two frames, using a range of different
   displacements in the radial distance, to find the maximum of the correlation coefficient. Our approach ensures a \mbox{sub-pixel} precision of the shift
   determination even for diffuse features, but it does not return the radial distance itself. Using the radial
   increments for all the pairs of {\em residual images} from each night and the radial distance of the chronologically
   first position from a crude estimation, we computed (using LSQ approach) the radial distances of the succeeding
   positions of a given CN structure. The retrieved positions are presented in Fig.~\ref{Threef} together with the
   expansion velocities projected onto the sky plane, obtained from a linear fit.

   In general, the projected velocities of different features are much smaller than the expansion velocity
   of 1.0 \mbox{km s$^{-1}$}, considered as canonical for the gas coma at 1 AU (e.g. Combi et al. \cite{combi04}).
   For November 2, 4 and 5 they range between 0.1 and 0.3 \mbox{km s$^{-1}$}, except for arc {\bf A} on November 2,
   which exhibited movement toward the coma centre with the velocity of 0.23 \mbox{km s$^{-1}$}.
   This unusual arc does not correlate with any other arc- or \mbox{spiral-like} feature visible in Fig.~\ref{Onef}.
   As the nature of this phenomenon remains unlear at the moment, we leave it for the subsequent model analysis.   

   As has been shown by Waniak et al. (\cite{waniak09}), CN produced by photodissociation of  HCN is a good tracer
   of its parent for the radial distances up to about one photochemical scalelength of HCN. Since the maximum radial
   distance considered in our analysis (\mbox{$\sim$2$\times$10$^{-4}$} km) is only twice the scalelength for HCN
   (\mbox{$\sim$10$^{-4}$} km), the condition is reasonably satisfied. Hence, the CN expansion velocity has to correlate well
   with the HCN expansion velocity. Obviously, it is hard to imagine that the expansion of the gas coma in
   103P was indeed as slow as the projected expansion of the CN structures. The projection effect would be negligible
   only if we observed expansion of a spherical shell or if the structures were exactly in the
   \mbox{sky-plane}. The real expansion velocity was most probably significantly reduced
   by the \mbox{on-sky} projection indicating mass loss into a cone with a given opening angle and orientation.
   The November 6 data
   revealed a much higher expansion velocity of 0.66 \mbox{km s$^{-1}$} for arc {\bf A} and 0.43 \mbox{km s$^{-1}$}
   for arc {\bf B}, which can be possibly
   explained by a significantly changed geometry of the gas ejection, resulting in a strong reduction of the
   projection effect.

\section{Periodicity in the CN data}

   The CN {\em residual images} were used as an input time series to search for periodicity. We separately
   investigated ({\em i}) variability in the photometric signal, and ({\em ii}) repeatability of the transient features.

   First, we performed aperture photometry of the central region of the CN coma, using a diaphragm with
   the optimum, \mbox{12-arcsec} diameter (increasing the diaphragm size would result
   in a higher level of undesirable smoothness and phase delay of the near nucleus signal
   variability, but decreasing the diameter would cause a higher noise pollution of the
   photometric data and greater influence
   of imperfect \mbox{on-comet} tracking, variable seeing, and defocus). We checked that the diaphragm
   twice as small changes the behaviour of the photometric data negligibly. That is because the expected aperture
   effect is stifled by the phase delay introduced by the photodissociation
   of the CN parent.
   The timescale for the aperture effect (i.e., the travel time between the aperture centre and its edge) is of the order
   of \mbox{10$^{3}$ s} and the typical photodissociation timescale for the CN parent
   at \mbox{1 AU} heliocentric distance is of the order of \mbox{10$^{4}$ s}.

\begin{figure}
  \resizebox{\hsize}{!}{\includegraphics{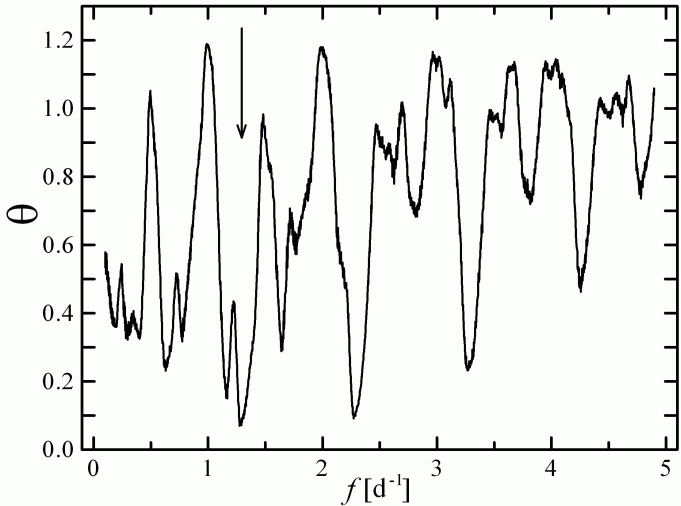}}
  \caption{PDM periodogram for our photometric data calculated using 6 bins and 6 covers.
           The optimal solution is found at the frequency of 1.310 \mbox{d$^{-1}$}
           (period of \mbox{18.32 h}) and is pointed by an arrow.}
  \label{Fourf}
\end{figure}

   The periodicity was investigated using the weighted version of the Phase Dispersion Minimization (PDM), developed by Drahus \& Waniak (\cite{drahus06}) from the classical
   \mbox{non-weighted} PDM (Stellingwerf \cite{stellingwerf78}). Quality of data phasing is probed by a parameter $\theta$,
   which is a ratio of variances of the phased and unphased data. The resulting PDM periodogram is presented in Fig.~\ref{Fourf}.
   The global minimum of $\theta$ occurs at the periodicity of \mbox{18.32$\pm$0.30 h}. Other minima are produced
   by the interference with a \mbox{one-day} cycle or represent harmonics of the basic frequency. Our result is an instantaneus synodic
   rotation period for the epoch UT 2010 Nov. 5.0835, i.e. for the middle moment of our run.
   Compared to other results obtained for similar dates, our solution is the same as the \mbox{18.32$\pm$0.03 h}
   periodicity detected in the HCN data (Drahus et al. \cite{drahus11}), agrees within the errors with the periods
   of \mbox{18.34$\pm$0.04 h} from the {\em EPOXI} photometry (A'Hearn et al. \cite{ahearn11}) and
   \mbox{18.195$\pm$0.010 h} obtained from radar Doppler imaging (Harmon et al. \cite{harmon11a}), and is similar
   to \mbox{18.7$\pm$0.3 h} (Knight \& Schleicher \cite{knight11}) and \mbox{$\sim$18.8 h}
   (Samarasinha et al. \cite{samarasinha11}) inferred from the repeatability of the CN coma
   profile.
\begin{figure}
  \resizebox{\hsize}{!}{\includegraphics{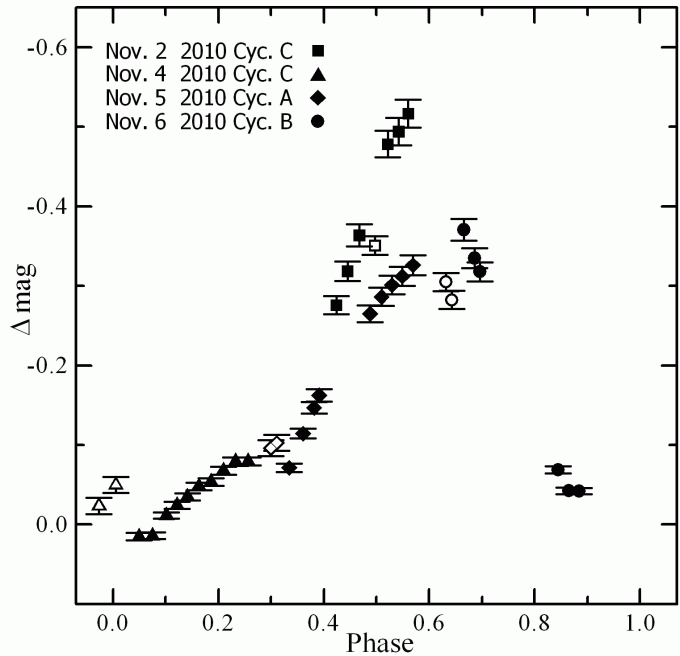}}
  \caption{Molecular light curves for CN (filled symbols) and C$_3$ (open symbols) phased
           according to the \mbox{18.32 h} period. The moment of the {\em EPOXI} encounter has a phase 0.5 of
           {\em Cycle B}.}
  \label{Fivef}
\end{figure}

   The light curve phased according to the period of \mbox{18.32 h} is displayed in Fig.~\ref{Fivef},
   where we show a magnitude difference between the {\em residual frames} and the
   \mbox{time-invariant} average coma profile produced by the {\em Iterative Image Decomposition}. Since the latter
   can
   be contaminated by dust (in contrast to the {\em residual images}; cf. Section 2), the presented light curve
   can be slightly deformed. For comparison, the C$_3$ photometric data are also presented in this plot.
   Although the C$_3$ signal could be contaminated by dust
   more strongly than CN, the photometric variability is comparable for both radicals.
   The maxima of the CN signal
   visible close to phase 0.5 can be linked with periodical insolation of the smaller tip of the elongated
   nucleus, that was active during the {\em EPOXI} encounter. Figure \ref{Fivef} shows that the maximum levels of the CN signal
   most probably differ noticeably for {\em Cycles} {\em A}, {\em B}, and {\em C}. The secondary increase of the CN
   signal at phases \mbox{$\sim$0.25} of {\em Cycle C} (November 4) is related to the weak central enhancement
   (see Fig.~\ref{Onef}) and could be attributed to less active vents. 

\begin{figure}
  \resizebox{\hsize}{!}{\includegraphics{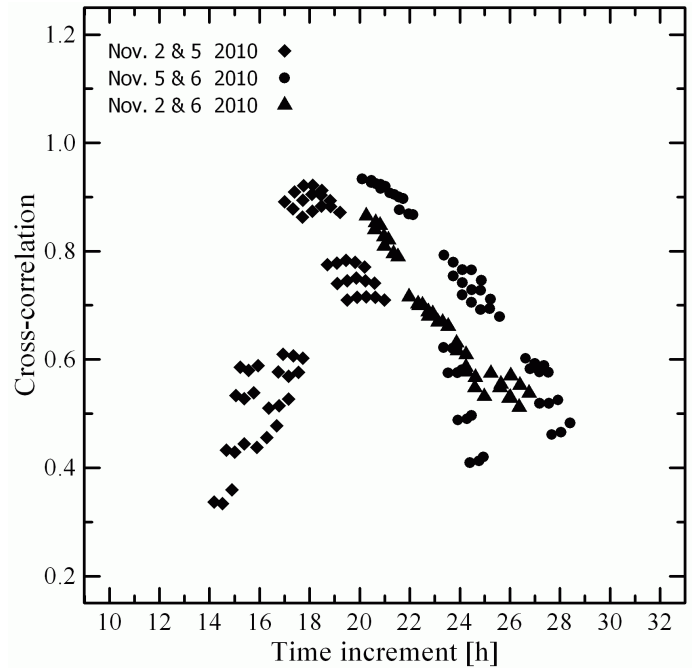}}
  \caption{Cross-correlation of the central CN shells between three nights.
           Multiples of the \mbox{18.32 h} period have been subtracted from the actual time
           increments to obtain differences closest to the single period.}
  \label{Sixf}
\end{figure}

   Another approach to investigate the periodicity in our {\em residual images} of CN is analysis of repeatability
   of the CN transient structures based on \mbox{cross-correlation} of the patterns in the {\em residual frames}. It is
   independent of the signal and measures the real similarity of the profiles. Although this procedure uses
   a different parameter than we used in the earlier work on comet 8P/Tuttle, the philosophy behind did not
   change, and can be found in Waniak et al. (\cite{waniak09}). Because the outer CN structures, detached from the central
   features, present diversity of profiles and expansion velocities, we have compared only the central features
   visible on November 2, 5 and 6. The resulting set of the \mbox{cross-correlation} parameters related to the
   time increments between two exposures is presented in Fig.~\ref{Sixf}.
   Unfortunately, none of the dependencies characterising
   the three pairs of nights present a clear maximum that could be used to precisely measure
   the period. While the plot
   is not as convincing as the photometric result in Fig.~\ref{Fivef}, it nevertheless provides an independent confirmation
   of our solution. Worth noticing are the relatively high levels of \mbox{cross-correlation}, reaching up to 0.95 for
   November 2 and 5, and November 5 and 6. Even though the central shells observed on November 2, 5 and 6 look
   differently at first sight (see Fig.~\ref{Onef}), the quantitative analysis shows that in general the patterns
   are similar.

\section{Differences between the nucleus precession cycles}

   First, we compare the behaviour of the CN coma in three consecutive precession cycles: {\em Cycle A}, {\em B},
   and {\em C}, which we observed during the \mbox{production-rate} maxima on November 2
   ({\em Cycle C}), 5 ({\em Cycle A}) and 6 ({\em Cycle B}). After analysing the behaviour of the central CN shells
   in the {\em residual images} (Fig.~\ref{Onef}), and taking into account
   that we showed (Fig.~\ref{Sixf}, Section 5)
   a higher correlation level for the pairs of nights November 2 and 5, and November 5 and 6 than for November 2
   and 6, we assumed that on November 5 ({\em Cycle A}) we observed the most common and least structured pattern,
   related presumably to the most quiescent precession cycle. Taking this \mbox{CN-shell} profile from November 5
   ({\em Cycle A}) as a reference, we monitored how the central shells for November 2 ({\em Cycle C}) and November
   6 ({\em Cycle B}) differ from
   it and from each other. First, we stacked the last four {\em residual images} from November 5 in order to increase the
   \mbox{signal-to-noise} ratio. Next we subtracted this {\em reference pattern} from the {\em residual images}
   for November 2 and 6.
\begin{figure}
  \centering
  \includegraphics[width=84mm]{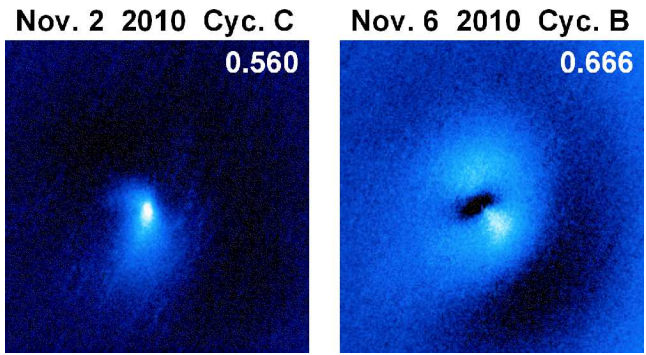}
  \caption{Comparison of the transient structures in the CN coma on November 2
          ({\em Cycle C}) and November 6 ({\em Cycle B}). Each panel presents
          the result of subtraction of the {\em reference residual image} (November 5)
          from the actual {\em residual frame}. Phases are displayed in the upper-right corners.
          Orientation and image scale are identical as in Fig.~\ref{Onef}.}
  \label{Sevenf}
\end{figure}

   Figure \ref{Sevenf} presents one example of the subtraction result for November 2 and one for November 6. The
   striking difference is easily visible.
   While on November 2 a bright jet directed southward was present, the November 6
   image shows an elliptical envelope with a pair of bright spots. The jet visible on the first night was brightening
   with time and then diminished a bit. It contributed to the CN signal in our aperture, increasing it to the highest
   level observed during the whole run. The night of November 6 ({\em Cycle B}) provides us with
   the CN pattern that closely resembles the CN structure from November 4 ({\em Cycle C}) but is about two times
   smaller. The position angles of the bright spots from November 6 correlate very well with the position angles
   of the similar features from November 4.
   Discrepancies are \mbox{1\degr} and \mbox{10\degr} for structures {\bf A} and {\bf B} respectively, but anyway
   of the order of the error bars. The reason for this similarity is that we observe different evolution stages
   (phase difference \mbox{$\sim$0.4}) of the same structure which appears periodically every three precession
   cycles.
   More precisely, these are the similarities between two different {\em Cycles B}: one
   on November 6 and the other one preceding {\em Cycle C} on November 4.
   We note, however, that both {\em Cycles B} exhibited different profiles of the CN features
   and dissimilar evolution. Moreover, the CN structures visible on November 4 moved slowly (0.14 \mbox{km s$^{-1}$}
   for part {\bf A} and 0.28 \mbox{km s$^{-1}$}  for feature {\bf B}), whereas the structures for November 6 expanded
   much faster (0.66 \mbox{km s$^{-1}$}  and 0.43 \mbox{km s$^{-1}$}  for structures {\bf A} and {\bf B} respectively).
   Taking into account the very stable radial evolution of the corkscrew structure on November 6 (especially its
   \mbox{south-west} part), not exhibiting the spiral or wavy pattern, we suspect that this phenomenon was created by
   a fan with a relatively large opening angle enclosing the Earthward direction. The fan was rotating as the
   nucleus was rotating about the precession axis which was enclosed in this fan as well. The source of such a flat feature having a large extent
   could be a properly oriented deep trench. Such topographic element could lie along the border
   between the ''waist`` region and one of the nucleus lobes.

   To explain the unexpected rise of the expansion velocity for this corkscrew pattern compared to the slowly
   moving CN features on the previous nights we need a significantly different influence of the projection effect.
   If our interpretation of the corkscrew structure is true,
   the \mbox{on-sky} projection did not affect markedly the observed velocity on November 6 and hence it should be
   close to the real expansion velocity of the gas. This possibility implies that the opening angle of the fan
   producing the
   corkscrew structure was indeed huge compared to the opening angle of the CN ejection on November 4.

   Comet 103P presents evident dissimilarities in the CN transient patterns and their
   \mbox{on-sky} kinematics (see Section 4) for successive {\em Cycles} {\em A}, {\em B} and {\em C}.
   Similar variations in the \mbox{line-of-sight} kinematcs of HCN, the most likely progenitor of CN, have been
   reported by Drahus et al. (\cite{drahus12}). The most natural explanation seems to be the one with a typical
   gas flow velocity of the order of 1 \mbox{km s$^{-1}$} and variation in the direction of the dominant
   gas ejection which changes markedly from one cycle to another.
   The \mbox{precession-roll} scenario, proposed by A'Hearn et al. (\cite{ahearn11}), appears to provide a
   plausible explanation and relates well to the extraordinary \mbox{spin-down} of the comet's nucleus resulting
   most probably from the jet effect, which can also enforce the excited mode of rotation
   (e.g. Guti\'errez et al. \cite{gutierrez03}).
   In such a case, one and the same active vent (e.g. the region recorded by {\em EPOXI} on the
   smaller tip of the nucleus) can eject matter in different directions and also other vents can be insolated
   during some or all of the precession cycles. 
\begin{figure} 
  \centering
  \includegraphics[width=84mm]{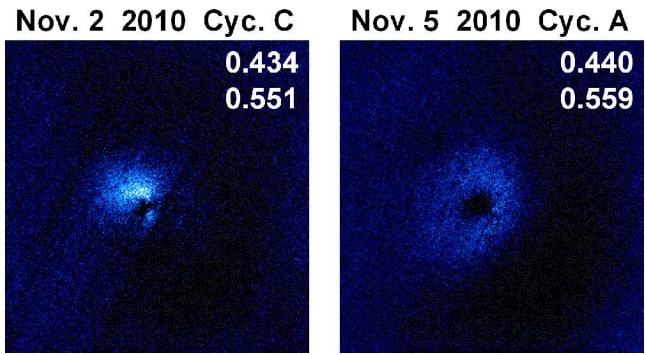}
  \caption{Evolution of the central CN shell observed on November 2 ({\em Cycle C})
           and November 5 ({\em Cycle A}).
           The behaviour in time is displayed by the result of subtraction of the earlier
           exposure from the later one. Precession phases are shown in the upper right corners.
           Orientation and image scale are the same as in Fig.~\ref{Onef}.}
  \label{Eightf}
\end{figure}

   Since the three consecutive precession cycles produced different coma structures, it is interesting
   to examine if these structures also evolved differently. Analysing video clip Animation2 we conclude that
   the expansion of the central shell observed on November 2 ({\em Cycle C}) varied substantially compared to the
   behaviours observed on November 5 ({\em Cycle A}) and 6 ({\em Cycle B}). The former case presents an expanding,
   \mbox{spiral-like} structure in the northern part
   of the central feature. To ensure that this is not an artefact produced by our version of the \mbox{{\em ring-masking}}
   procedure (used to generate video clip Animation2) we verified the result using a simple
   algebra for the {\em residual images}. To enhance the \mbox{signal-to-noise} ratio we stacked image \#1 with \#2, and \#4 with
   \#5 from November 2, and also \#4 with \#5, and \#8 with \#9 from November 5. Consequently, for each of the two
   nights we obtained a pair of images with the time differences equal to 2.1 and 2.2 hours for
   November 2 and November 5, respectively. Then we subtracted the earlier image
   from the later one in each pair (all the frames were carefully normalized before). Such differential images
   (Fig.~\ref{Eightf}) clearly show the evolution of the central CN patterns as the difference of signals correlates with
   time increment, i.e. higher values imply later stages of evolution. The results confirm
   our previous findings.
   On November 2 we observed the rotating \mbox{spiral-like} structure produced by a clockwise
   nucleus precession as projected on the sky. This case resembles the well developed spirals reported for 103P in
   \mbox{September-October} (see e.g. Samarasinha et al. \cite{samarasinha11}). The ellipsoidal shell
   observed on November 5 expanded highly symmetrically. Such a behavior is fully consistent with
   the scenario of gas ejection into a wide cone enclosing the Earthward direction.
   Prolateness of this envelope is controlled by the nucleus precession that changes the orientation of the active
   tip during the insolation phase producing the typical spiral which is visible almost completely \mbox{edge-on} on this
   date. We observe a similar situation on November 6.
   
   The dissimilarity in the evolution of the central CN shell noticed between {\em Cycles C} and {\em A} may be
   caused by the roll about the longest axis of the nucleus (A'Hearn et al. \cite{ahearn11}). It is sufficient
   that the active vent, which occupies the smaller tip of the nucleus, is shifted off the roll axis.
   
\section{Orientation of the nucleus rotation axis}

   We determine the orientation of the rotation axis assuming for simplicity that the nucleus was a SAM rotator
   with the spin axis which is close to the precession axis. We noticed in Section 6 that the corkscrew structure
   (especially its \mbox{south-west} part) visible on November 6 expanded strictly radially and
   the position angles of its expansion direction correlate with the traces of the corkscrew pattern from November 4.
   Hence, for different precession cycles there is one distinguished direction, which we attribute to the \mbox{on-sky}
   projection of the rotation axis of 103P. If so, the \mbox{south-west} section of the corkscrew structure was
   created by a circumpolar gas jet. As the \mbox{on-sky} velocity of this pattern was relatively high, it
   expanded in a direction far from the line of sight, forming a conical structure consistent with
   our observations. Thus, we have assumed that the symetry axis of the \mbox{south-west} part
   of the corkscrew structure corresponds with the \mbox{on-sky} projection of the nucleus rotation axis.
 
\begin{figure}
  \resizebox{\hsize}{!}{\includegraphics{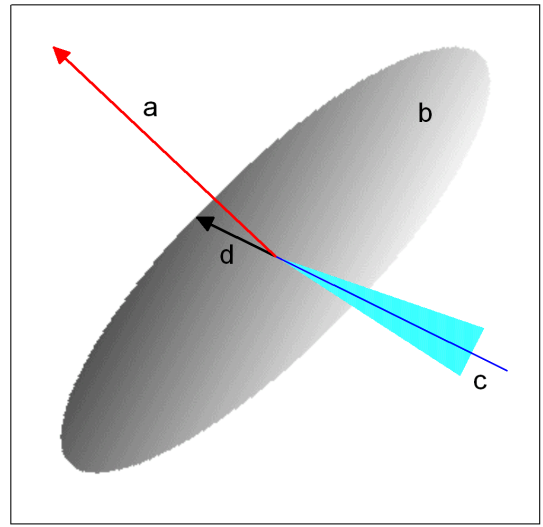}}
  \caption{Sketch illustrating how the orientation of the rotation axis was determined. This is an
           \mbox{on-sky} projection with orientation like in Fig. 1. Circular plane {\bf\em b} is perpendicular to
           vector {\bf\em a}, which represents the direction of the shortest axis of inertia of the comet's nucleus.
           Line segment {\bf\em c} shows the \mbox{on-sky} direction of the axis of the corkscrew pattern.
           Angular momentum vector {\bf\em d} shows our solution. It should be located at plane {\bf\em b}, and its
           \mbox{on-sky} projection should coincide with line segment {\bf\em c}.}
  \label{Tenf}
\end{figure}

   In the next step we used the {\em in situ} determination of an instantaneous orientation of
   the longest axis
   of 103P's nucleus during the {\em EPOXI} flyby (A'Hearn et al. \cite{ahearn11}), which we consider as representing
   the shortest axis of
   inertia. It is obvious that the rotation axis has to lie in the plane perpendicular to this axis.
   Figure \ref{Tenf} presents this situation. On the other hand, the \mbox{on-sky} projection of the
   rotation axis should coincide with the direction of the \mbox{south-west} part of the corkscrew pattern.
   Combining both restrictions we find the orientation of the north rotation pole (the sense of spinning is determined
   in Section 6) as \mbox{RA=122\degr} and \mbox{Dec=+16\degr} (epoch J2000.0).
   The error of our determinantion is dominated by the validity level of our assumptions which is very
   difficult to estimate. If the assumptions were strictly satisfied, the obtained circular error would be \mbox{2\degr}
   which we computed by propagating the uncertainty of the position angle of the \mbox{south-west} part of the corkscrew
   structure and the error of the orientation of the longest nucleus axis (A'Hearn et al. \cite{ahearn11}).

   A'Hearn et al. (\cite{ahearn11}) found the direction of the total angular momentum (close to the precession axis)
   equal to \mbox{RA=17$\pm$11\degr} and \mbox{Dec=+47$\pm$2\degr}.
   Radar Doppler imaging (Harmon et al. \cite{harmon11b}) gave the result \mbox{RA=332\degr} and \mbox{Dec=+20\degr}.
   Analysis of the CN coma structures made by Knight \& Schleicher (\cite{knight11}) shows
   the axis direction of \mbox{RA=257\degr} and \mbox{Dec=+67\degr} with the uncertainty of \mbox{15\degr}, and
   Samarasinha et al. (\cite{samarasinha11}) obtained \mbox{RA=345\degr}, \mbox{Dec=$-$15\degr} with a
   \mbox{20\degr} precision.
   Although, it is hard to obtain a coherent picture from these results, all the solutions (including
   ours) are preliminary at this stage.

\section{Comparison of CN and HCN}

\begin{figure}
  \resizebox{\hsize}{!}{\includegraphics{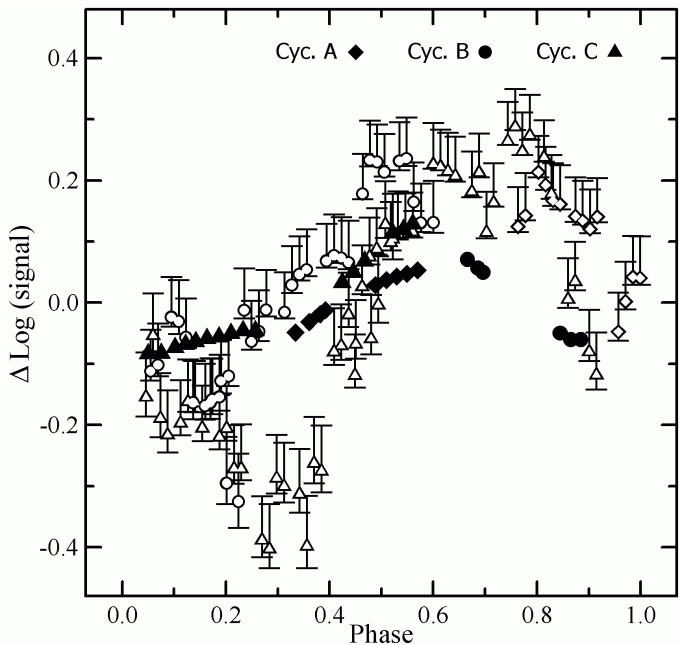}}
  \caption{Comparison between the molecular light curves of CN (see text)
           and HCN (from Drahus et al. \cite{drahus11}). The HCN signal is represented by the
           line area measured in the beam of \mbox{8.8-arcsec} FWHM. The light curves
           are calculated with respect to the mean levels, and phased according to the \mbox{18.32 h} period.
           The filled symbols are used for CN and the open symbols for HCN, while the symbol shape
           indicates the \mbox{{\em three-cycle}} component. The error bars for CN are of the order
           of the symbol size and have been omitted.}
  \label{Ninef}
\end{figure}
   Since we noticed the exceptional coincidence of the rotation periods from
   the HCN and CN data (cf. Section 5),
   we also examined the correlation between both molecules, comparing the variability in the relative photometry
   of CN (Section 5) with a series of the HCN \mbox{main-beam} brightness temperatures from IRAM \mbox{30-m}
   (the latter being a part of the observing material obtained by Drahus et al. \cite{drahus11}; \cite{drahus12}).
   Both the CN photometry and the HCN data selected
   for comparison were collected around the same time and partly simultaneously. Figure \ref{Ninef}
   presents this comparison,
   where the data are phased in the same way as before (cf. Sections 3 and 5), and identically
   for all practical purposes with Drahus et al. (\cite{drahus11}).

   It should be born in mind that HCN was monitored with the beam size three times smaller than the aperture
   used for our CN photometry. Hence, the CN variability can be \mbox{phase-delayed} as compared
   with HCN. Such an aperture effect was discussed in Section 5 and found negligible.
   The markedly smaller amplitude of variability for CN may result from smoothing introduced by the finite lifetime
   of the CN parent (\mbox{$\sim$10$^{4}$ s} for HCN). In general, both light curves correlate quite well in the active
   phases and much worse during the quiescent periods. Perhaps CN in 103P originates from two different sources, each
   behaving differently. One source could be related to insolated active vents and the other one to different
   parts of the nucleus surface or to a volume source surrounding the nucleus, such as e.g. the ''snowflakes`` detected
   by {\em EPOXI} (A'Hearn et al. \cite{ahearn11}). To look deeper into the correlation between HCN and CN
   a \mbox{non-steady-state} model of the \mbox{HCN-CN} coma has to be build, which is a subject of our ongoing
   work.

\section{Conclusions}

   \begin{enumerate}
      \item Image processing of the CN and C$_3$ frames revealed transient structures in the gas coma of 103P.
            Both radicals formed very similar features at close time
            moments. CN and C$_3$ photometry shows general consistence. We conclude that the progenitors of CN
            and C$_3$ were emitted from the same regions of the nucleus, which implies homogeneity of the
            CN and C$_3$ parents.
      \item The CN signal varied with a period of \mbox{18.32$\pm$0.30 h} at the epoch of our
            observations (UT 2010 Nov. 5.0835),
            the same as was measured for HCN around the same time. The variability of HCN
            correlates with the behaviour of CN suggesting that HCN was a major donor of CN.   
      \item The CN structures expanded with projected velocities between 0.1 and 0.3 \mbox{km s$^{-1}$}, except
            for November 6, when the velocity increased up to 0.66 \mbox{km s$^{-1}$} in the \mbox{south-west} part of
            the corkscrew pattern.
      \item The CN shells, appearing in the central part of the coma at the phases of boosted CN production, present
            a high level of \mbox{cross-correlation}. They were most probably generated
            by periodic insolation of the smaller tip of the nucleus (active at the {\em EPOXI} encounter).
            The CN profiles at the consecutive {\em Cycles}
            {\em A}, {\em B} and {\em C} differed markedly. A modest CN boosting was observed
            during {\em Cycle A} (November 5), a corkscrew pattern was present during {\em Cycle B} (November 6),
            and a southern jet was visible at {\em Cycle C} (November 2). They suggest that for different 
            cycles different nucleus parts, carrying different vents, were exposed to the solar radiation and active.
      \item Although the corkscrew structure seems to appear every three precession cycles, the situation did
            not repeat strictly (different profiles and kinematics).
      \item Combining the results of our image analysis with the instantaneous nucleus orientation from {\em EPOXI}
            and assuming the \mbox{principal-axis} rotation mode we obtained the orientation of the
            spin axis equal to
            \mbox{RA=122\degr} and \mbox{Dec=+16\degr} (epoch J2000.0).
   \end{enumerate}

\begin{acknowledgements}
     The authors gratefully acknowledge observing grant support from the Institute of Astronomy and Rozhen
     National Astronomical Observatory, Bulgarian Academy of Sciences. W. Waniak acknowledges the financial
     support from the Nicolas Copernicus Foundation for Polish Astronomy. M. Drahus was supported by a 
     NASA Planetary Astronomy grant to David Jewitt. The authors thank the referee, M. Belton, for very helpfull 
     suggestions.     
\end{acknowledgements}

\Online
\begin{appendix} 

 \section{Image cleaning procedure}

\begin{figure}
  \resizebox{\hsize}{!}{\includegraphics{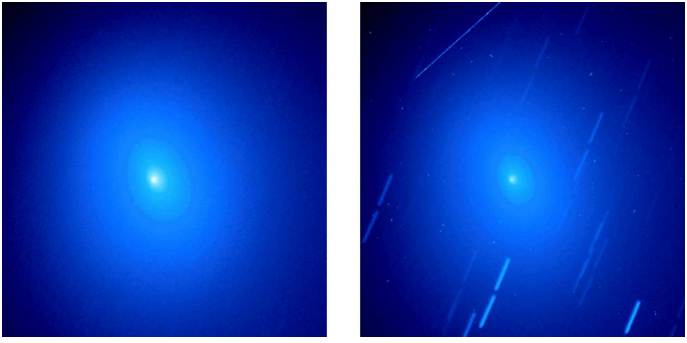}}
  \caption{Result of removal of CRHs and stellar profiles by the cleaning procedure (left pannel). 
           For comparison, the right panel shows the result of simple stacking of the same two frames.
           Date of observation is UT 2010 Nov. 7.1375. The field of view is 6.4 arcmin.
           North is up and east is to the left.}
  \label{Elevf}
\end{figure}

   With the aim to remove Cosmic Ray Hits (CRH) and stellar profiles elongated by \mbox{non-sidereal} tracking
   we grouped consecutive images in pairs and compared the first frame with the second frame in each pair.
   We took advantage of the statistical independence of CRH, and the fact that in two consecutive frames
   a given stellar trail occupies separable regions (the \mbox{40-s} CCD readout time corresponds to 3 arcsec
   of comet's motion).
   We subtracted and averaged  images in pairs, obtaining two series of images: one with differential frames
   and other one with mean frames. For each differential image we correlated its noise statistics with the
   mean signal in the corresponding averaged frame. Using the dependence between noise level and mean signal
   it is possible, for a given pixel, to estimate the dispersion of signal in the differential frame.
   When the pixel value in the differential image was higher than $K\sigma$ (i.e. a factor of $K$ above the local noise)
   we flagged it in an auxiliary binary mask frame. As the stellar patterns in the auxiliary mask frames contain
   regions with disconnected pixels (external envelopes of bright stars and whole patterns for stars at the
   noise limit) we produced compact patterns improving pixel connectivity. We used the \mbox{closing-like} procedure
   based on the local surface density of the flagged pixels. For a given pixel, the density was probed in a small
   circle surrounding this pixel and compared with the limiting value. This value was determined using the 
   \mbox{trial-and-error} procedure. Above this limit the pixel was kept flagged and vice versa. This stage of image
   processing was completed by synthesizing clean stacks. In the regions where no CRHs or stellar trails
   were detected, the mean values from both frames were taken. Otherwise, the values from the image with
   no detection were accepted. As the comet was observed in relatively dense stellar fields, the profiles
   of different stars occasionally overlapped in the images from one pair. In such situations the stellar
   trails were masked and filled using interpolation. By stacking two consecutive images we increased the
   \mbox{signal-to-noise} ratio except for the regions occupied by the stellar profiles or CRHs. We note that in
   rare cases one and the same frame was used to create two consecutive pairs. Obviously, such pairs are not
   independent, but let us use the entire nightly material, even if the number of single images is odd. Figure
   \ref{Elevf} presents one example of how our cleaning procedure is able to remove unwanted objects from the comet 
   image. 

\end{appendix}

\end{document}